\documentclass[review]{elsarticle}

\usepackage{hyperref}
\usepackage{amsfonts}
\usepackage{graphicx}
\usepackage{xcolor}
\usepackage{float}
\newtheorem{Teo}{Theorem}[section]
\newtheorem{Pro}{Proof}
\newtheorem{Lem}{Lemma}[section]
\newtheorem{Def}{Definition}[section]
\newtheorem{remark}{Remark}

\usepackage{amsmath}
\usepackage{amssymb}

\journal{Physic A}

\begin{document}

\begin{frontmatter}

\title{Katugampola Generalized Conformal Derivative Approach to Inada Conditions and Solow-Swan Economic Growth Model}
%\tnoteref{mytitlenote}

%\author{G. Fernandez-Anaya\fnref{myfootnote}}
%\address{Universidad Iberoamericana}

\author[mymainaddress]{G. Fern\'{a}ndez-Anaya}

\author[mysecondaryaddress]{L. A. Quezada-T\'{e}llez\corref{mycorrespondingauthor}}
\cortext[mycorrespondingauthor]{Corresponding author}
\ead{lquezada@correo.cua.uam.mx}

\author[mymainaddress]{B. Nu\~nez-Zavala}
\author[mymainaddress]{D. Brun-Battistini}

\address[mymainaddress]{Universidad Iberoamericana, Ciudad de Mexico, Mexico.}
\address[mysecondaryaddress]{Universidad Autonoma Metropolitana-Unidad Cuajimalpa, Ciudad de Mexico, Mexico.}

\begin{abstract}
This article shows a new focus of mathematic analysis for the Solow-Swan economic growth model, using the generalized conformal derivative Katugampola (KGCD). For this, under the same Solow-Swan model assumptions, the Inada conditions are extended, which, for the new model shown here, depend on the order of the KGCD. This order plays an important role in the speed of convergence of the closed solutions obtained with this derivative for capital ($k$) and for per-capita production ($y$) in the cases without migration and with negative migration. Our approach to the model with the KGCD adds a new parameter to the Solow-Swan model, the order of the KGCD and not a new state variable. In addition, we propose several possible economic interpretations for that parameter.
\end{abstract}


\end{frontmatter}

\section{Introduction}

The economic growth models of \cite{Solow1} and \cite{Swan}, that henceforth, because of the similarities between them, we will call the Solow-Swan model (SSM), are models that try to explain income and its growth in an economy through the amounts of resources involved in production (capital and labor) and technological change or technical progress. The model assumes, among other things, perfect competition, full employment, decreasing marginal returns in the use of capital and labor, constant returns to scale for a production function that is homogeneous and is considered exogenous because it includes variables or parameters whose value is determined outside the model or is an external data.

In the SSM, the technical progress, denoted by the constant $A$, is an exogenous parameter defined as a factor of increasing scale that multiplies the production function \cite{Solow1}  and includes in that concept the improvements in human factors through time \cite{Solow2}. Recently, the SSM continues to be studied, for example in  \cite{Grassetti} it is done using a Kadiyala production function.

Two are the main criticisms of SSM from the perspective of endogenous growth models: 1) its theoretical inability to explain long-term economic growth, which can only be achieved if it is imposed exogenously through technical progress (a strong technological change) to increase income and per-capita capital levels, as well as welfare of families over time, either before reaching their steady state (in which economy growth stops) or once this has been reached (\cite{Lucas} and \cite{Barro}), and 2) the impossibility of empirically verifying in the long term, the convergence that the model predicts for all the economies of the planet, those developed and those that are not, to the same steady state \cite{Barro}.

This second criticism has also been addressed by those \cite{Mankiw} who, based on the theories of endogenous economic growth, model the differences between the economies of the world, adding human capital (the knowledge, skills and competencies of workers as individuals) to the two state variables of the (SSM). When this variable was included, the convergence rates of the different types of economies to the same steady state improved, however, they remained insufficient.

Nonetheless, in all the mentioned cases and in others as case \cite{Juchem} that analyze the closed solution for the economic growth in relation to the migration, the method to obtain the solutions for the variables of the proposed models (endogenous and exogenous) was typical of derivatives of an integer order.

Other works using models based on fractional order derivatives have also addressed the problem of economic growth. For example, in \cite{Tarasova1}, a generalization of the economic model of natural growth is suggested, which takes into account the memory effect of the power law type. The memory effect implies the dependence of the process, not only on its current state but also on its history of changes in the past. For the mathematical description of the economic process with power law memory, the theory of non-integer derivatives and fractional differential equations is used. They conclude that the memory effect can lead to a decrease in the output and not to its growth, which is typical of a model without memory.

In \cite{Tarasova2}, economic models must take into account the memory effects caused by the fact that economic agents, in their decisions, remember the history of changes in the exogenous and endogenous variables characterizing economic processes. The continuous-time description of the economic processes by decreasing memory of the power law type can be described using fractional calculus and fractional differential equations. The inclusion of memory effects in economic models can lead to new results with the same parameters.

In \cite{Stamova} with sufficient conditions to analyze the Mittag Leffler stability type, a Solow type model of fractional order is introduced as a new tool in mathematical finance. The main advantage of the proposed approach is the non-locality property of these fractional derivatives that are convenient for the modeling of real financial situations and macroeconomic systems.

Models based on other types of derivatives have also been studied, for example in \cite{Brestovanska}, axioms for the system of differential equations (time scales) of Solow type are formulated under the assumption that a certain function is constant, proving stability and balance results in positive coordinates. A Cobb-Douglas type production function is also considered.

In \cite{Bohner}, a general Solow model on time scales is introduced from which a first-order nonlinear dynamic equation that describes the model is deduced. Based on a Cobb-Douglas type production function, several cases are considered when there is no technological development or changes in the population. Subsequently, they consider the case with technological development and population growth. These models include, as particular cases, different types of derivatives such as quantum calculus and some of conformal type.

The case of slow growth when the order of the fractional derivative $\rho \in (0,1)$  is similar to the situation that appears in physics with the description of anomalous diffusion models \cite{Metzler}, which is related in form of sub-diffusion for $0<\rho<1$. The anomalous diffusion equations have been used to describe financial processes \cite{Scalas,Mainardi,Gorenflo,Laskin,Cartea,Mendes,Machado}.

We propose derivatives of conformal type \cite{Khalil,Katugampola} that could, without increasing state variables, offer an alternative to solve the two criticisms raised previously against the SSM for the case with negative migration. The model obtained with the KGCD is simpler than the models presented with fractional derivatives, or with derivatives of the time-scales type. In addition, we present closed solutions of the SSM with the KGCD, very similar to the model of the integer SSM order, and therefore it is possible to give an analogous interpretation to those of the classical model.

Applying the KGCD of order $\rho$ \cite{Katugampola} instead of the usual derivative of the integer order in the SSM, we show two results that preserve the Inada conditions, imposing restrictions on the $\rho$ order of the KGCD related to the marginal contribution of capital ($\alpha$) to the production obtained ($Y$). Later, we obtain closed solutions for capital and per-capita production on time for zero and negative migration cases. These closed solutions with the KGCD introduce a new parameter $\rho$, and not a new state variable to the solutions obtained for the SSM as other authors do. The analysis of the SSM model with the KGCD allows us to show that the $\rho$ parameter plays a fundamental role in maintaining consistency with Inada's well-known conditions. As expected, when the parameter $\rho$ takes the value of $1$, we recover the derivative of integer order, and consequently the classic SSM.

This article is organized as follows. In the second section the mathematical preliminaries are presented. In the third section the congruence of the KGCD with the Inada conditions of integer order is shown. In the fourth section, closed solutions are obtained for the SSM model without migration and with negative migration. In the negative migration case, the analysis on the restrictions imposed for capital and per-capita production is presented, and also the times for which the aforementioned variables increase to infinity. In the sixth section some representative graphs of the results obtained in the previous sections are presented. Finally, the general conclusions are presented.

\section{Preliminaries of KGCD}

This section provides the main definitions of the KGCD, as well as the foundation of the SSM. 

\subsection{Basic Definitions on KGCD}

This subsection discusses the definition of the conformal derivative, as well as its properties of order $\rho \in [0,1]$.

\begin{Def}  \cite{Katugampola}
Given a function $f:[0,\infty) \rightarrow \mathbb{R}$. Then the KGCD of order $\rho$, $m$ is defined by:
\begin{equation}
D_{m}^{\rho}f(t)={lim}_{\epsilon\rightarrow 0 }\frac{f(te_{k}^{\epsilon t^{-\rho}})-f(t)}{\epsilon},
\label{EqDef1_1}
\end{equation}
where
\begin{equation}
e_{m}^{t}=\sum_{i=0}^{m}\frac{t^{i}}{i!},
\label{EqDef1_2}
\end{equation}
for all $t>0,\rho \in [0,1]$.
\label{Def1}
\end{Def}

As a consequence of the previous definition, the following Lemma is obtained.

\begin{Lem}
If $f$ is differentiable, then $D_{m}^{\rho} f(t)=t^{1- \rho} \frac{df}{dt}$.
\label{Lema1}
\end{Lem}

\begin{Pro}
Let's take 
\begin{eqnarray}
h(t,\epsilon)= & \epsilon t^{1-\rho}\left(1+\epsilon t^{1-\rho}+\frac{\epsilon^{2}t^{2\left(1-\rho\right)}}{2!}+...+\epsilon^{m}\frac{t^{m\left(1-\rho\right)}}{m!}\right) \nonumber \\
=& \epsilon t^{1-\rho}\left(1+O_{m}(\epsilon)\right)=\epsilon t^{1-\rho}+\hat{O}_{m}(\epsilon^{2}),
\label{EqPro1}
\end{eqnarray}
where $\hat{O}_{m}(\epsilon^{2})=\epsilon t^{1-\rho}O_{m}\left(\epsilon\right)$. Then, the definition of the KGCD of order $\rho$, $m$ is as follows:

\begin{equation}
D_{m}^{\rho} f(t)={lim}_{\epsilon \rightarrow 0} \frac{f(te_{m}^{\epsilon t^{-\rho}})-f(t)}{\epsilon}=
\label{EqPro2}
\end{equation}

\begin{equation}
{lim}_{\epsilon \rightarrow 0} \frac{f(t+{\epsilon t^{1-\rho}}+\hat{O}_{m}(\epsilon^{2}))-f(t)}{\epsilon}=
\label{EqPro3}
\end{equation}

\begin{equation}
{lim}_{\epsilon \rightarrow 0} \frac{f(t+h(t,\epsilon))-f(t)}{\epsilon}=
\label{EqPro4}
\end{equation}

\begin{equation}
{lim}_{\epsilon \rightarrow 0} \frac{f\left(t+h\left(t,\epsilon\right)\right)-f(t)}{\frac{h\left(t,\epsilon\right)t^{\rho-1}}{1+O_{m}\left(\epsilon\right)}}=
\label{EqPro5}
\end{equation}

\begin{equation}
{lim}_{\epsilon \rightarrow 0} t^{1-\rho} \frac{df(t)}{dt}.\
\label{EqPro6}
\end{equation}
	
If the KGCD of $f$ exists, then we will say that the function $f$ is $\rho-$differentiable in some interval $(0,a)$ with $a>0$, the KGCD of the fuction $f$ exits and the ${lim}_{t\rightarrow0^{+}}D_{m}^{\rho} f(t)$ exists as well, hence we define
	
\begin{equation}
D_{m}^{\rho} f(0)={lim}_{t\rightarrow0}D_{m}^{\rho} f(t).
\label{EqPro7}
\end{equation}
	
Note that
	
\begin{equation}
D_{1}^{\rho} f(t)={lim}_{\epsilon \rightarrow 0} \frac{f(t+\epsilon t^{1-\rho})-f(t)}{\epsilon}
\label{EqPro8}
\end{equation}
is the KGCD defined for $m=1$ and
	
\begin{equation}
D_{\infty}^{\rho} f(t)={lim}_{\epsilon \rightarrow 0} \frac{f(te^{\epsilon t^{-\rho}})-f(t)}{\epsilon}
\label{EqPro9}
\end{equation}
is the KGCD defined for $m=\infty$. \qed
\label{Pro1}
\end{Pro}

Let it be noted that the derivatives (\ref{EqPro8}) and (\ref{EqPro9}) are particular cases of the derivative KGCD (\ref{EqDef1_1}). Which correspond to the conformal derivative in \cite{Khalil} and the derivative of Katugampola in \cite{Katugampola}. Therefore, the following results are similar due to Lemma (\ref{Lema1}). 

\begin{Teo} \cite{Khalil,Katugampola}
If a function $f:[0,\infty) \rightarrow \mathbb{R}$ is $\rho-$differentiable in $t_{0}>0,\rho \in (0,1]$, so $f$ is continuous on $t_{0}$.  
\label{Teorema1}
\end{Teo}

\begin{Teo} \cite{Khalil,Katugampola}
Let it be that $\rho \in (0,1]$ and $f,g$ are $\rho-$differentiable for $t>0$. Therefore
\begin{enumerate}[1)]
\item $D_{m}^\rho(af+bg)=aT_{\rho}(f)+bT_{\rho}(g)$ for every $a,b \in \mathbb{R}$. 
\item $D_{m}^\rho(t^{q})=qt^{q-\rho}$ for all $q \in \mathbb{R}$.
\item $D_{m}^\rho(\lambda)=0$, for all constant functions $f(t)=\lambda$. 
\item $D_{m}^\rho(fg)=fT_{\rho}(g)+gT_{\rho}(f)$.
\item $D_{m}^\rho\left(\frac{f}{g}\right)=\frac{gT_{\rho}(f)-fT_{\rho}(g)}{g^{2}}$. 
\label{Teorema2}
\end{enumerate}
\end{Teo}

In the following section, the Inada conditions are presented applying the KGCD.

\section{Inada conditions for the Solow-Swan model with the KGCD}

The SSM is a benchmark for most economic growth analysis. In time, the model is represented as in Eq.(\ref{FuncProd})  where $Y$ is the national production, $K$ and $L$ (state variables) are the quantities of capital and labor factors used in production, both measured with the appropriate units and $A$ represents a technological constant that is usually interpreted as the total productivity of all factors. For a period of time $t$, national production is the result of combining capital and labor, given a certain technological constant, represented mathematically as:

\begin{equation}
Y(t)=F(A,K(t),L(t))
\label{FuncProd}
\end{equation}
where $t$ indicates time \cite{Juchem}. 

\begin{Def}
The production $Y(t): R^{+} \rightarrow R$ is a function and must fulfill the following properties \citep{Juchem}, known in Economic Science as the Inada conditions. Replacing the derivative of integer order with the KGCD, we obtain the following modified Inada conditions:
\begin{enumerate}[i)]
\item The function $Y(t)$ is increasing for both state variables, capital ${(D_{K,m}^\rho Y>0)}$ and labour force $(D_{L,m}^\rho Y>0)$.
\item The function $Y(t)$ will have constant returns to scale, ${Y(\lambda K,\lambda L)=\lambda Y(K,L)}$, $\forall \lambda>0$. 
\item The function $Y(t)$ satisfies the conditions: ${\displaystyle\lim_{K \rightarrow 0}D_{K,m}^\rho Y =\displaystyle\lim_{L \rightarrow 0}D_{L,m}^\rho Y = +\infty}$ and $\displaystyle\lim_{K \rightarrow \infty}D_{K,m}^\rho Y =\displaystyle\lim_{L \rightarrow \infty}D_{L,m}^\rho Y = 0$
\item The function $Y(t)$ also satisfies that $(D_{K,m}^{2\rho} Y<0)$ and that $(D_{L,m}^{2\rho} Y<0)$
\end{enumerate}
Where:
$D_{K,m}^\rho Y$ y $D_{L,m}^\rho Y$ are the partial derivatives of $Y$ of order $\rho$, $m$ of KGCD with respect to the variables $K$ and $L$ respectively.
\label{Def2}
\end{Def}

Let it be noted that when $\rho$ is $1$ we recover the usual Inada conditions of integer order. Thereafter, we present the following propositions, where we will consider $A,K,L > 0$. Let's observe the following:
\begin{equation}
D_{K,m}^\rho Y(K,L) = K^{1-\rho} D_{K,m} Y(K,L)
\label{EqProp1}
\end{equation}

\begin{equation}
D_{L,m}^\rho Y(K,L) = L^{1-\rho} D_{L,m} Y(K,L)
\label{EqProp2}
\end{equation}

\begin{align}
D_{K,m}^{2\rho} Y(K,L) &=D_{K,m}^\rho (D_{K,m}^\rho Y(K,L))  \nonumber \\
D_{K,m}^{2\rho} Y(K,L) &=K^{1-2\rho}[(1-\rho)D_{K,m}Y + KD_{K,m}^2 Y]
\label{EqProp3}
\end{align}

and
\begin{align}
D_{L,m}^{2\rho} Y(K,L) &=D_{L,m}^\rho (D_{L,m}^\rho Y(K,L)) \nonumber \\
D_{L,m}^{2\rho} Y(K,L) &=L^{1-2\rho}[(1-\rho)D_{L,m}Y + LD_{L,m}^2 Y]
\label{EqProp4}
\end{align}

\begin{Teo}
If the Inada conditions for the KGCD are satisfied for some ${\rho \in (0,1)}$, then the Inada conditions of integer order are satisfied.
\label{Teorema3}
\end{Teo}

\begin{remark}
Let it be noted that we do not assume an explicit form for the function  $Y(K,L)$.
\end{remark}

\begin{Pro} 
Let's suppose that $\rho_{0} \in (0,1)$ exists so that the Inada conditions are satisfied with the KGCD (\ref{Def2}). Therefore
\begin{enumerate}[i)]
	\item Since $D_{K,m}^{\rho_{0}} Y =  K^{1-\rho} D_{K,m}Y > 0$, then $D_{K,m}Y > 0$ and similarly \\ $D_{L,m}^{\rho_{0}} Y= L^{1-\rho} D_{L,m}Y > 0$, then $D_{L,m}Y > 0$.
	\item Is the same statement as for $\rho = 1$.
	\item If $\lim_{K \rightarrow 0} D_{K,m}^{\rho_{0}} Y = \lim_{K \rightarrow 0} K^{1-\rho_{0}} D_{K,m} Y = \infty$, then $D_{K,m} Y$ it is of order $O(K^{-(1-\rho_{0}+\epsilon)})$ with $\epsilon > 0$. Therefore $\lim_{K \rightarrow 0}D_{K,m}Y = \infty$. Similarly, $\lim_{L \rightarrow 0} D_{L,m}^{\rho_{0}} Y = \infty$ implies $\lim_{L \rightarrow 0} D_{L,m} Y = \infty$.
	
	If $\lim_{K \rightarrow \infty} D_{K,m}^{\rho_{0}} Y = \lim_{K \rightarrow \infty} K^{1-\rho_{0}} D_{K,m} Y = 0$, then $D_{K,m}Y$ is of order $O(K^{-(1-\rho_{0}+\epsilon)})$ with $\epsilon > 0$. Therefore $\lim_{K \rightarrow \infty} D_{K,m} Y = 0$. The case of $\lim_{L \rightarrow \infty} D_{L,m} Y = 0$ is similar. 
	\item $D_{K,m}^{2\rho_{0}} Y < 0$ if and only if $K^{1-2\rho_{0}}[(1-\rho_{0})D_{K,m}Y + KD_{K,m}^2 Y]<0$, whereby $(1-\rho_{0})D_{K,m}Y + KD_{K,m}^2 Y <0$, but $(1-\rho_{0})$, $K$ and $D_{K,m}Y$ are positive because of item i), therefore $D_{K,m}^2 Y <0$. Similarly $D_{L,m}^{2\rho_{0}} Y < 0$, then $D_{L,m}^2 Y < 0$. \qed
\end{enumerate}
\label{Pro2}
\end{Pro}

\begin{Teo}
Let's consider the SSM, with production function $Y=AK^\alpha L^{1-\alpha}$. The Inada conditions for the KGCD are satisfied if and only if $\rho > max [\alpha, 1-\alpha]$.
\label{Teorema4}
\end{Teo}

\begin{Pro}
First let's note that $Y=AK^\alpha L^{1-\alpha}$ satisfies the Inada condition of integer order.
\begin{enumerate}[i)]
	\item $D_{K,m}^\rho (Y) = K^{1-\rho} D_{K,m}(Y) > 0$ and $D_{L,m}^\rho (Y) = L^{1-\rho} D_{L,m}(Y) > 0$ because $K,L,D_{K,m}(Y)$ y $D_{L,m}(Y)$ are positive for $Y=AK^\alpha L^{1-\alpha}$.
	\item It is the same statement for integer order and KGCD.
	\item $\lim_{K \rightarrow 0} D_{K,m}^\rho(Y) = \lim_{K \rightarrow 0}  K^{1-\rho} D_{K,m}(AK^\alpha L^{1-\alpha})=$ \\
	$ \alpha AL^{1-\alpha} \lim_{K \rightarrow 0}  K^{\alpha -\rho} = \infty$, if $\rho>\alpha$. \\
	$\lim_{L \rightarrow 0} D_{L,m}^\rho(Y)=\lim_{L \rightarrow 0}L^{1 -\rho} D_{L,m}(Y) = $ \\
	$AK^\alpha (1-\alpha) \lim_{L \rightarrow 0} L^{1-(\rho +\alpha)} = \infty$, if $\rho > 1- \alpha$. Therefore ${\rho > max[\alpha,1-\alpha].}$ \\
	Analogously $\lim_{K \rightarrow \infty}D_{K,m}^\rho (Y)= \alpha AL^{1 -\rho} \lim_{K \rightarrow \infty} K^{\alpha-\rho} = 0$, if $\rho > \alpha$.  Similarly $\lim_{K \rightarrow \infty}D_{K,m}^\rho (Y)= 0$, if $\rho > 1-\alpha$. Therefore $\rho > max[\alpha,1-\alpha]$.
	\item  $D_{K,m}^{2\rho} (Y) = K^{1-2\rho}[(1-\rho)D_{K,m}(Y)+KD_{K}^2(Y)]<0$ if and only if 
\begin{equation}
|D_{K}^2 (Y)|> \frac{1-\rho}{K}|D_{K,m}(Y)|
\label{Eq17}
\end{equation}

because $D_{K,m}(Y) > 0$ y $D_{K,m}^2(Y) < 0$. 

	Substituting $Y=AK^\alpha L^{(1-\alpha)}$ in Eq. (\ref{Eq17}) and considering the absolute values, we obtain $A\alpha(1-\alpha)K^{\alpha-2}L^{1-\alpha}>(1-\rho)A\alpha K^{\alpha -2} L^{1-\alpha}$ if and only if $1- \alpha > 1 - \rho$ equivalently $\rho > \alpha$.
	Similarly $D_{L,m}^{2\rho} (Y) < 0$ if $|D_{L,m}^2 (Y)| > \frac{(1-\rho)}{L}|D_{L,m}(Y)|$ $\Longleftrightarrow$ $AK^\alpha(1-\alpha)L^{-\alpha-1}(1-\rho) < AK^\alpha (1-\alpha) \alpha L^{-\alpha-1}$ given that $1-\rho < \alpha $ $\Longleftrightarrow$ $\rho > 1-\alpha$, therefore both conditions are fulfilled $\Longleftrightarrow$ $\rho> max[\alpha,1-\alpha]$.
	Notice that $\alpha \in [0,1]$ implies that $\rho \in (1/2, 1]$ since $\rho$ cannot be greater than $1$ by the propositions that consider the Cobb-Douglas form $Y=AK^\alpha(t) L^{1-\alpha}(t)$ , $\alpha \in (0,1)$. \qed
\end{enumerate}
\label{Pro3}
\end{Pro}

These results show that the Inada conditions of integer order are preserved with the new derivative KGCD. In the next section the SSM is described using the KGCD.

\section{The Solow-Swan model with the KGCD}

In this section we analyse two cases: the case without migration with a Malthusian law considering the population as labour force and the case with negative constant migration. We will follow a similar procedure to the one developed in \cite{Juchem}. 
If the production function is of the Cobb-Douglas form \cite{Solow1}:

\begin{equation}
Y=AK^{\alpha}(t)L(t)^{1-\alpha},  \hspace{0.5cm}  \alpha \in (0,1)
\label{CobbDouglas}
\end{equation}
when $\alpha$ is close to $0$ it is said that the economy is work intensive and for the opposite case, if $\alpha$ is almost $1$, it is capital intensive. According to (\ref{CobbDouglas}), the capital stock dynamics is governed by the ordinary differential equation:

\begin{equation}
\dot{K}=sY-\delta K=sAK^{\alpha}L^{1-\alpha}-\delta K
\label{stockcapital}
\end{equation}
where $s$ and $\delta$ are the savings constants and the rate of depreciation of capital respectively, hence, neoclassically, $sY$ can be taken as the gross investment and $\delta  K$ is the capital depreciation of the entire economy \cite{Juchem}. In the following subsections we present both cases, with and without migration.

\subsection{The KGCD Solow Growth Model without Migration}

Deriving out of the SSM of integer order, we present in this subsection a new model, applying the KGCD. In it, the KGCD retrieves the properties of integer order, but introduces a parameter in the order of the derivative that allows greater flexibility to the model and is compatible with the classic SSM when $\rho$ takes the unit value.

The rest of the work will use the following notations indistinctly: \\
${L_{m}^{(\rho)} = D_{m}^\rho (L)}$

\begin{equation}
L^{(\rho)}_{m}=\gamma L\Rightarrow L(t)=L_{0}e^{\gamma\frac{t^{\rho}}{\rho}}
\label{MalthsinMigra}
\end{equation}
where $L_{0}>0$ is the initial population of workers and $\gamma$ the inter-temporal rate of growth or Malthusian parameter. If we replace (\ref{MalthsinMigra}) in (\ref{CobbDouglas}) we obtain

\begin{equation}
Y=AK^{\alpha}\left(L_{0}e^{\gamma\frac{t^{\rho}}{\rho}}\right)^{1-\alpha}
\end{equation}

If we define the per-capita capital, as
\begin{equation}
k(t)=\frac{K(t)}{L(t)}
\label{kpercapita}
\end{equation}

and the labour growth rate as, 

\begin{equation}
n(t)=\frac{L_{m}^{(\rho)}(t)}{L(t)}
\label{tasacreclab}
\end{equation}

Therefore, deriving (\ref{kpercapita}) with respect to the time and taking $n(t)=\gamma$, we obtain that $k_{m}^{(\rho)}=\frac{K_{m}^{(\rho)}}{L}-\gamma k(t)$. From (\ref{stockcapital}) and replacing now the capital stock using the KGCD is given by 

\begin{align}
K_{m}^{(\rho)}=sY-\delta K &=sAK^{\alpha}L^{1-\alpha}-\delta K \nonumber \\
k_{m}^{(\rho)}+(\delta+\gamma)k &=sAk^{\alpha}
\label{BernoulliFrac}
\end{align}
where (\ref{BernoulliFrac}) is a Bernoulli equation using the KGCD. Through Bernoulli's well-known variable change, to obtain the linear equation, we take $w=k^{1-\alpha}$,  

\begin{align}
w_{m}^{(\rho)} &=(1-\alpha)k^{-\alpha}k^{(\rho)} \nonumber \\
w_{m}^{(\rho)}+(1-\alpha)(\gamma+\delta)w &=(1-\alpha)sA
\label{operadointeg}
\end{align}

From (\ref{BernoulliFrac}) and (\ref{operadointeg}) the solution for $k(t)$ is given by 

\begin{equation}
k(t)=\left[c_{1}e^{-(1-\alpha)(\gamma+\delta)\frac{t^{\rho}}{\rho}}+\frac{sA}{\gamma+\delta}\right]^{\frac{1}{1-\alpha}}
\label{Soluc1}
\end{equation}
where $c_{1}$ is a constant of appropriate units.

It is important to note that the steady state of per-capita capital $k_{\infty}$, is given by  

\begin{equation}
k_{\infty}=\displaystyle\lim_{t \rightarrow +\infty} k(t)=\left(\frac{sA}{\gamma + \delta}\right)^\frac{1}{1-\alpha}
\label{CapPerCap}
\end{equation}

Note that the limit, when $t$ tends to infinity, coincides with the limit of integer order. On the other hand, we now define the per-capita product as the total production ratio in respect to work, that is,

\begin{equation}
y(t)=\frac{Y(t)}{L(t)}=Ak^{\alpha}(t)
\label{CapPerCap1}
\end{equation}
where using the expression (\ref{CobbDouglas}), in the long term the per-capita production tends to

\begin{equation}
y_{\infty}=\displaystyle\lim_{t \rightarrow +\infty} Ak^{\alpha}(t)=A\left(\frac{sA}{\gamma+\delta}\right)^\frac{1}{1-\alpha}.
\label{ProdPerCap}
\end{equation}
The solutions obtained here with KGCD have the same mathematical behavior as those of classical SSM. Therefore in the case without migration, the convergence to the same steady state of capital and per-capita production will exist regardless of the value of $\rho$. However, in this case the convergence speed decreases as the value of $\rho$ decreases.

\subsection{The KGC Solow Growth Model with Migration}

Under the same assumptions, this subsection presents another KGCD model adding a constant migration rate $(I)$ to the differential equation (\ref{MalthsinMigra}), which determines the labour force of the economy, therefore we have:

\begin{equation}
L_{m}^{(\rho)}=\gamma L+I.
\label{MalthsconMigra}
\end{equation}

Calculating the solution of (\ref{MalthsconMigra})

\begin{equation}
L(t)=ce^{\gamma\frac{(t-t_{0})^{\rho}}{\rho}}+\int_{0}^{t}e^{\gamma\frac{(t-t_{0})^{\rho}}{\rho}}e^{-\gamma\frac{(s-t_{0})^{\rho}}{\rho}}I(s-t_{0})^{\rho-1}ds
\label{MalthsconMigra2}
\end{equation}

where $t_{0}$ is an initial time which we take with value $0$ in  (\ref{MalthsconMigra2}), we obtain

\begin{equation}
L(t)=\left[\left(c+\frac{I}{\gamma}\right)e^{\gamma\frac{t^{\rho}}{\rho}}-\frac{I}{\gamma}\right], \hspace{0.5cm} c=L_{0}
\label{L}
\end{equation}

Note that if $\rho=1$, we retrieve the solution of the integer migration case. Therefore, if a labour growth rate is taken as (\ref{tasacreclab}) and the KGCD is applied, it results in

\begin{equation}
\bar{n}(t)=\frac{L_{m}^{(\rho)}}{L}=\frac{\gamma(\gamma L+I)}{(\gamma L_{0}+I)e^{\gamma\frac{t^{\rho}}{\rho}}-I}
\label{tasacremigralab}
\end{equation}

Again, with (\ref{stockcapital}) and now replacing the stock capital with migration using the KGCD, the per-capita capital satisfies the following equation

\begin{equation}
\bar{k}_{m}^{(\rho)}+(\bar{n}(t)+\delta)\bar{k}=SA\bar{k}^{\alpha}
\label{stockcapmigrac}
\end{equation}
where (\ref{stockcapmigrac}) was obtained analogously to Eq.(\ref{BernoulliFrac}) but considering migration. This is also a Bernoulli equation with the KGCD. If we now take $Z=\bar{k}^{1-\alpha}$ and $Z(0)=Z_{0}=\bar{k}_{0}^{1-\alpha}$ with $Z_{m}^{(\rho)}=(1-\alpha)\bar{k}^{-\alpha}\bar{k}_{m}^{(\rho)}$, for a similar argument we come to:
\begin{equation}
Z_{m}^{(\rho)}-(\alpha-1)(\delta+\bar{n}(t))Z=(1-\alpha)sA.
\label{BernoulliMigra}
\end{equation}

Now using Eq.(\ref{tasacremigralab}), we obtain that:

\begin{equation}
\int\left(\delta+\bar{n}\left(t\right)\right)t^{\rho-1}dt=\delta\frac{t^{\rho}}{\rho}+\int\frac{d \gamma L(t)}{\gamma L(t)} =\delta\frac{t^{\rho}}{\rho}+ln\left[\left(\gamma L_{0}+I\right)e^{\gamma\frac{t^{\rho}}{\rho}}-I\right]-ln(\gamma L_{0})
\label{Ec36}
\end{equation}
where $\gamma L(t) = (\gamma L_{0} + I)e^{\frac{\gamma t^{\rho}}{\rho}} - I$. 

On the other hand, substituting the integral of the Eq.(\ref{Ec36}) in the exponent of the exponential function, we get:

\begin{eqnarray}
e^{\pm \int_{0}^{y}\left(  1-\alpha\right)  \left(  \delta+\bar{n}(t)\right)
t^{\rho-1}dt}=e^{\pm \left( 1-\alpha \right)  \delta\frac{y^{\rho}}{\rho}}\left[  \left(
\frac{\gamma L_{0}+I}{\gamma L_{0}}\right)  e^{\frac{\gamma y^{\rho}}{\rho}}-\frac{I}{\gamma L_{0}}\right]
^{\pm {\left ( 1- \alpha \right)}}.
\end{eqnarray}

Therefore the solution of Eq.(\ref{BernoulliMigra}) is:
\begin{align}
Z(t) &=Z_{0} e^{-\int\left(1-\alpha\right)(\delta+\bar{n}(t))t^{\rho-1}dt}+e^{-\int\left(1-\alpha\right)(\delta+\bar{n}(t))t^{\rho-1}dt}  \nonumber  \\
&\qquad {}  \int e^{\left( 1-\alpha \right)  \delta\frac{y^{\rho}}{\rho}}\left[  \left(
\frac{\gamma L_{0}+I}{\gamma L_{0}}\right)  e^{\frac{\gamma y^{\rho}}{\rho}}-\frac{I}{\gamma L_{0}}\right]
^{\left( 1- \alpha \right)} (1-\alpha)sAt^{\rho-1}dt
\label{SolucZeta1}
\end{align}

Hence, from the following equation it is possible to obtain the capital and per-capita production:

\begin{align}
Z(t) &=Z_{0}e^{\left(  \alpha-1\right)  \delta\frac{t^{\rho}}{\rho}}
[\left( \frac{ \gamma L_{0}+I}{\gamma L_{0}}\right)  e^{\frac{\gamma t^{\rho}}{\rho}}-\frac{I}{\gamma L_{0}}]^{\left(  \alpha-1\right)}
+ \nonumber \\
&\qquad {}  \left(  1-\alpha\right)  sAe^{\left(  \alpha-1\right)  \delta\frac{t^{\rho}}{\rho}} [\left( \frac{ \gamma L_{0}+I}{\gamma L_{0}}\right)  e^{\frac{\gamma t^{\rho}}{\rho}}-\frac{I}{\gamma L_{0}}]^{\left(  \alpha-1\right)} \nonumber \\
&\qquad {}   \left(\int_{0}^{t}\left(  e^{-\left(  \alpha-1\right)  \delta \frac{y^{\rho}}{\rho}} \left[  \left(  \frac{ \gamma L_{0}+I}{\gamma L_{0}}\right)  e^{\frac{\gamma y^{\rho}}{\rho}}-\frac{I}{\gamma L_{0}}\right]  ^{\left( 1- \alpha\right)} y^{\rho-1}dy \right)\right)
\label{SolucZeta2}
\end{align}
where it is obtained that the per-capita capital $(\bar{k})$ is given by
\begin{eqnarray}
\bar{k}(t)=Z(t)^{\frac{1}{1-\alpha}}
\label{SolucsMigk}
\end{eqnarray}
and per-capita production $(\bar{y})$ by
\begin{eqnarray}
\bar{y}(t)=AZ(t)^{\frac{\alpha}{1-\alpha}}.
\label{SolucsMigy}
\end{eqnarray}

The last two equations (\ref{SolucsMigk}), (\ref{SolucsMigy}), give us closed solutions for capital and per-capita production with migration. In the next section we will give the closed solution for the case of negative migration by explicitly solving the integral of (\ref{SolucZeta2}) in terms of hypergeometric functions.  

\subsection{The Closed Analytic Solutions of KGCD Solow Growth Model with Migration}

In this section we applied a similar procedure to used in \cite{Juchem} for to solve the integral of the second term of the equation (\ref{SolucZeta2}).

\begin{equation}
J=\int_{0}^{t} e^{(1-\alpha) \frac{\delta \tau^{\rho}}{\rho}} \left[ \left( \frac{\gamma L_{0} + I}{\gamma L_{0}}\right){e^{\frac{\gamma \tau^{\rho}}{\rho}}}-\frac{I}{\gamma L_{0}}\right]^{(1-\alpha)}\tau^{(\rho - 1)} d\tau
\end{equation}

Applying the change of variable $u=e^{\frac{\gamma t^\rho}{\rho}}$ we have that, \\ 
$du=\frac{\gamma}{\rho}e^{\frac{\gamma \tau^\rho}{\rho}}\left(\rho \tau^{(\rho - 1)} d\tau \right)=\gamma u \tau^{(\rho - 1)}d\tau$
from where $\frac{du}{\gamma u}=\tau^{(\rho - 1)} d\tau$

\begin{equation}
J=\frac{1}{\gamma}\int_{1}^{e^{\frac{\gamma \tau^{\rho}}{\rho}}} u^{\frac{(1-\alpha)\delta}{\gamma}-1}\left[-\frac{I}{\gamma L_{0}} +\left( \frac{\gamma L_{0} + I}{\gamma L_{0}}\right)u \right]^{(1-\alpha)} du
\end{equation}

\begin{equation}
J=\frac{1}{\gamma}\left(-\frac{I}{\gamma L_{0}} \right)^{(1-\alpha)} \int_{1}^{e^{\frac{\gamma t^{\rho}}{\rho}}} u^{\frac{(1-\alpha)\delta}{\gamma}-1}\left[1-\left(1+\frac{\gamma L_{0}}{I}\right )u \right]^{(1-\alpha)} du
\end{equation}

The last integral is related to Euler's integral representation of the Gaussian Hypergeometric Function $_{2}F_{1}$:

\begin{align}
_{2}F_{1}\left(  \left.
\begin{array}
[c]{c}%
a,b\\
c
\end{array}
\right\vert z_{1}\right) &=\sum_{n=0}^{\infty} \frac{(a)_{n}(b)_{n}z^n}{(c)_{n} n!}  \nonumber \\
&=\frac{\Gamma(c)}{\Gamma(c-b)\Gamma(b)}\int_{0}^{1}t^{b-1}(1-t)^{c-b-1}(1-zt)^{-a}dt
\end{align}
where $(.)_{n}=\Gamma ( . +n)/\Gamma( . )$ is the Pochhammer symbol \cite{Andrews,Erdelyi}. The series is convergent for any $a, b, c$ if $|z|< 1$, and for $Re({a+b-c})<0$ if $|z|=1$. For the integral representation $Re(c)>Re(b)>0$ is needed. In this case $\Gamma(z)$ denotes the Gamma Function. Thus,

\begin{equation}
J=\frac{1}{\gamma}\left(-\frac{I}{\gamma L_{0}} \right )^{(1-\alpha)} \left( J_{t}-J_{0} \right) 
\end{equation}

where
\begin{equation}
J_{0}=\frac{\gamma}{(1- \alpha)\delta} \hspace{0.1cm} _{2}F_{1}\left(  \left.
\begin{array}
[c]{c}%
a,b\\
c
\end{array}
\right\vert z_{1}\right)
\end{equation}

where
\begin{equation}
J_{t}=\frac{\gamma e^{(1-\alpha)\delta \frac{t^{\rho}}{\rho}}}{(1- \alpha)\delta}  \hspace{0.1cm} _{2}F_{1}\left(  \left.
\begin{array}
[c]{c}%
a,b\\
c
\end{array}
\right\vert z_{2}(t)\right)
\end{equation}

with $a=\alpha -1$, $b=\frac{(1-\alpha) \delta}{\gamma}$, $c=\frac{(1-\alpha) \delta}{\gamma}+1$, $z_{1}=\left(1+\frac{\gamma L_{0}}{I}\right)$ y $z_{2}(t)=\left(1+\frac{\gamma L_{0}}{I}\right)e^{\gamma \frac{t^{\rho}}{\rho}}$. And therefore, we obtain formulas that are explicitly closed for capital and per-capita production with negative migration

\begin{align}
Z(t) &= Z_{0}e^{\left(  \alpha-1\right)  \delta\frac{t^{\rho}}{\rho}}
[\left( \frac{ \gamma L_{0}+I}{\gamma L_{0}}\right)  e^{\frac{\gamma t^{\rho}}{\rho}}-\frac{I}{\gamma L_{0}}]^{\left(  \alpha-1\right)}
+ \nonumber \\
&\qquad {}  \left(  1-\alpha\right)  sAe^{\left(  \alpha-1\right)  \delta\frac{t^{\rho}}{\rho}} [\left( \frac{ \gamma L_{0}+I}{\gamma L_{0}}\right)  e^{\frac{\gamma t^{\rho}}{\rho}}-\frac{I}{\gamma L_{0}}]^{\left(  \alpha-1\right)} \nonumber \\
&\qquad {}   \left(\frac{1}{\gamma}\left(-\frac{I}{\gamma L_{0}} \right )^{(1-\alpha)} \left( J_{t}-J_{0} \right)  \right)
\end{align}

\begin{align}
\bar{k}(t) &= e^{-  \delta\frac{t^{\rho}}{\rho}} [\left( \frac{ \gamma L_{0}+I}{\gamma L_{0}}\right)  e^{\frac{\gamma t^{\rho}}{\rho}}-\frac{I}{\gamma L_{0}}]^{-1} \nonumber \\
& \left[  \bar{k}_{0}^{1-\alpha} +   \left(\frac{(1-\alpha)sA}{\gamma}\left(-\frac{I}{\gamma L_{0}} \right )^{(1-\alpha)} \left( J_{t}-J_{0} \right)  \right)  \right]^{\frac{1}{1-\alpha}} 
\label{SolMigNegK}
\end{align}

\begin{equation}
\bar{y}(t)=A \bar{k}(t)^{\alpha}
\label{SolMigNegY}
\end{equation}

The $\left(-\frac{I}{\gamma L_{0}} \right )^{(1-\alpha)} $ factor implies that: if $I \leq 0$, therefore $\gamma > 0$, and if $I > 0$, then $\gamma < 0$. In the next section we offer restrictions on the values that migration $I$ and time $t$ can take.

\section{Analysis for $I$ negative.}

In this section an analysis is made of the restrictions on capital and per-capita production, for $I<0$ and $\gamma>0$ according to the following lemmas.

\begin{Lem}
\begin{enumerate}[i)]
\item $L(t)=\left(  \frac{I}{\gamma}+L_{0}\right)  e^{\gamma\frac{t^{\rho}%
}{\rho}}-\frac{I}{\gamma}>0$ with $\gamma>0$, $0<\frac{I}{\gamma}+L_{0}$ and
$I<0,$ if $I\in\left[  -\gamma L_{0},0\right]  .$
\item $L(t)=\left(  \frac{I}{\gamma}+L_{0}\right)  e^{\gamma\frac{t^{\rho}%
}{\rho}}-\frac{I}{\gamma}>0$ with $\gamma>0$, $\frac{I}{\gamma}+L_{0}<0$ and
$I<0,$ if $I\in\left(  -\infty,-\gamma L_{0}\right)  $ y $t<t_{f}$ where
$t_{f}=\left[  \ln\left(  1+\frac{\gamma L_{0}}{I}\right)  \right]
^{\frac{-1}{\gamma}}.$
\end{enumerate}
\label{Lema5.1}
\end{Lem}

\begin{Pro}
\begin{enumerate}[i)]
\item If $\gamma>0$, $0<\frac{I}{\gamma}+L_{0}$ and $I<0\Leftrightarrow-\gamma
L_{0}<I<0\Leftrightarrow$ ${I\in\left[  -\gamma L_{0},0\right]}  \Rightarrow
L(t)=\left(  \frac{I}{\gamma}+L_{0}\right)  e^{\gamma\frac{t^{\rho}}{\rho}%
}-\frac{I}{\gamma}>0.$

\item If $\gamma>0$, $\frac{I}{\gamma}+L_{0}<0$ and $I<0\Leftrightarrow I<-\gamma L_{0}\Leftrightarrow I\in\left(  -\infty,-\gamma L_{0}\right)$ and $\left(
\frac{I}{\gamma}+L_{0}\right)  e^{\gamma\frac{t^{\rho}}{\rho}}>\frac{I}%
{\gamma}\Leftrightarrow$ $e^{\gamma\frac{t^{\rho}}{\rho}}>\frac{\frac
{I}{\gamma}}{\frac{I}{\gamma}+L_{0}}=\frac{1}{1+\frac{\gamma L_{0}}{I}%
}\Rightarrow t<t_{f}=\left[  \ln\left(  1+\frac{\gamma L_{0}}{I}\right)
\right]  ^{\frac{-1}{\gamma}}$. \qed
\end{enumerate}
\end{Pro}

\begin{Lem}
$\gamma>0$ and $\left\vert z_{1}\right\vert <1$ and $\left\vert
z_{2}\right\vert <1\Leftrightarrow I\in\left(  -\infty,-\gamma L_{0}\right)  $
y $t<t_{f}=\left[  \ln\left(  1+\frac{\gamma L_{0}}{I}\right)  \right]
^{\frac{-1}{\gamma}}.$
\label{Lema5.2}
\end{Lem}

\begin{Pro}
$\left\vert z_{1}\right\vert <1\Leftrightarrow-1<1+\frac{\gamma L_{0}}%
{I}<1\Leftrightarrow-2<\frac{\gamma L_{0}}{I}<0.$

$\left\vert z_{2}\right\vert <1\Leftrightarrow-1<\left(  1+\frac{\gamma L_{0}%
}{I}\right)  e^{\gamma\frac{t^{\rho}}{\rho}}<1\Leftrightarrow\frac{-1}%
{1+\frac{\gamma L_{0}}{I}}<e^{\gamma\frac{t^{\rho}}{\rho}}<\frac{1}%
{1+\frac{\gamma L_{0}}{I}}$

$\frac{-1}{1+\frac{\gamma L_{0}}{I}}<e^{\gamma\frac{t^{\rho}}{\rho}}$ is fulfilled if $1+\frac{\gamma L_{0}}{I}>0\Leftrightarrow-1<\frac{\gamma L_{0}}{I}
$.

If $\gamma>0$ and $I<0$ or $\gamma<0$ and $I>0$ then $\frac{-1}%
{1+\frac{\gamma L_{0}}{I}}<1=e^{0}\Rightarrow{-2<\frac{\gamma L_{0}}{I}<0}.$

Note that $-1<\frac{\gamma L_{0}}{I}<0\Leftrightarrow I\in\left(
-\infty,-\gamma L_{0}\right)$. On the another hand, ${e^{\gamma\frac{t^{\rho}%
}{\rho}}<\frac{1}{1+\frac{\gamma L_{0}}{I}}}$ is fulfilled if $t<t_{f}$. Therefore, $\gamma>0$ and $\left\vert z_{1}\right\vert <1$ and $\left\vert
z_{2}\right\vert <1$ $\Leftrightarrow I\in\left(  -\infty,-\gamma
L_{0}\right)$ and $t<t_{f}$. \qed
\end{Pro}

\begin{remark}
If the conditions of lemma  \ref{Lema5.2} are met, so are the conditions of item ii) of the lemma \ref{Lema5.1}. Hence, we can only consider the case $I\in\left(  -\infty,-\gamma L_{0}\right)$ and $t<t_{f}$.
\end{remark}

As a consequence of the previous results, it is not possible to perform an asymptotic analytic for $\bar{k}(t)$ when $t\rightarrow\infty$ and for any $\rho\in\lbrack0,1]$, because the hypergeometric functions $J_{t}$ y $J_{0}$ converge only for $\left\vert z_{1}\right\vert <1$ and $\left\vert z_{2}\right\vert <1$ which implies that $I\in\left(
-\infty,-\gamma L_{0}\right)$ and $t<t_{f}$. However, we can calculate the limit $\lim_{t\rightarrow t_{f}}\bar{k}(t)=\infty$, which coincides with the case of integer order. 

Note that $L(t)> 0$ is a hypothesis for the Inada conditions to be met with the KGCD and in the case of an integer order, and for the function of Cobb-Douglas production to take real values. Therefore, the only possible case is the one we have just determined.

In general, when applying the KGCD to the obtained solutions for capital and per-capita production, a new parameter $\rho$ is incorporated, and not a new state variable to the solutions obtained for the SSM model. This parameter $\rho$ represents the order of the KGCD applied and can vary between $0$ and $1$ to meet the Inada conditions if the KGCD is applied. It allows to recover the solutions of the SSM when $\rho$ adopts its maximum value and is consistent with the Inada conditions, and therefore with the SSM, if its value is greater than $max[\alpha,1-\alpha]$.      

In a simple way, the parameter $\rho$ added to the classic SSM when using the KGCD, can be interpreted as the speed with which an economy approaches towards its steady state in which there will be no further growth. The aforesaid speed will be slower the smaller the said parameter is. In this trajectory, per-capita capital decreases and, therefore, per-capita production and the economy's growth rate do it as well. 

In this sense, the possibility of incorporating into the classic SSM a set of differences between the economies of the planet that makes their trajectories towards the same steady state do not coincide in speed and/or value. Therefore, there is the possibility of solving the second criticism of said model that was aforementioned. 

The $\rho$ parameter could, in addition, allow economic science researchers to model endogenously (and without adding a state variable) some data of variables from the economic, socio-political and institutional spheres that, the closer they are to zero, more they decelerate the falling rate of the trajectories of capital and the per-capita production over time towards its respective long-term steady state.

Among the useful economic data to represent $\rho$ there could be:  
1) The percentage of available resources not used, that is, the percentage of idle capacity of available resources. 2) The percentage of profits (payment to capital) that is used for purposes other than the creation of new capital through innovation. 3) The inflation rate of an economy, this is the speed with which prices increase. At a higher rate, per-capita capital and, therefore, also per-capita production will fall to lower values and more rapidly than in the case of economies with lower inflation rates. 4) The interest rate. 5) The tax rate applied in direct and indirect taxes. 6) The unemployment rate. 7) The natural rate of unemployment. 8) The size of the total public deficit with respect to GDP. 9) The size of imports regarding GDP. 10) The complement of the ratio of the Trade Openness Index $(1-TOI)$. 11) The standardized risks measured between zero and one by the rating agencies. 12) The percentage of concentration in the economic structure (monopolies and oligopolies). 13) The degree to which other market failures arise (externalities, incomplete markets, public goods and social goods, among others). In the socio-political and institutional spheres, $\rho$ could represent, among other things, the degrees of: 14) Risk perceived by economic agents. 15) Informality in the economy. 16) Inequality in income distribution. 17) Poverty and Malnutrition. Health and education shortfalls (illiteracy). 18) Corruption. 19) Impunity. 20) Social violence. 21) Public insecurity. 22) The efficiency with which private economic agents, with their rational expectations, inhibit countercyclical economic policy measures that the government designs and executes to reduce and decelerate the capital and per-capita production downfalls.  

One more possibility is that $\rho$ is an index that results from a combination of all or some of the aspects of any of the areas indicated previously, provided it reflects in an aggregated and hierarchical manner the differences between the economies of the planet in normalized values between zero and one.

Finally, $\rho$ could also represent the rate of diminishing returns of capital that generates convergence to the same steady state and that could be different for each economy of the world, since it would be determined by the economic, socio-political and institutional aspects listed above.

\section{Graphics of some representative examples }

In this section we present the different cases of the solution for capital and for production, both per-capita, without migration and with negative migration. In Figures 1 and 2 the following values were taken: $\gamma=0.02$, $\alpha=0.6$, $\delta=0.05$, $s=0.12$, $A=1$, $k_{0}=200$, $L_{0}=100$, and $-\gamma L_{0}=-2$. For the Figures 3 and 4 $\gamma=0.14$, $\alpha=0.69$, $\delta=0.19$, $s=0.19$, $A=1$, $k_{0}=100$, $L_{0}=100$, and $H=-19.0$. 

\subsection{Case without migration}

\begin{figure}[ht]
\centering
\includegraphics[width=0.6\textwidth]{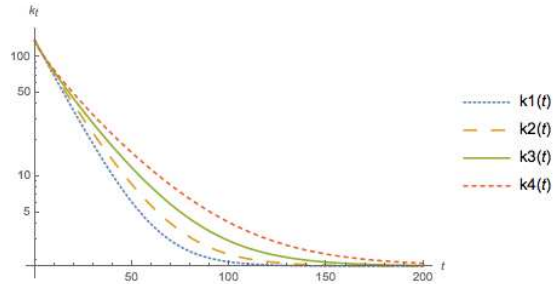}
\caption{The value of per-capita capital $k1(t)$ without migration of integer order $\rho=1.0$, $k2(t)$ with fractional order of $\rho=0.95$, $k3(t)$ with fractional order of $\rho=0.90$ and $k4(t)$ with fractional order of $\rho=0.85$.}.  
\label{Fig1}
\end{figure}

Figure (\ref{Fig1}) shows the trajectories of per-capita capital for the case without migration and for different values of $\rho >max[\alpha,1-\alpha]$ compared with the one of integer order $(\rho = 1)$. It can be noted that, without migration $(I=0)$ and with a $s$ lower than $\alpha$, the per-capita capital decreases at a lower speed while the order of the derivative KGCD is smaller and it does so converging to the same steady state regardless of the value of $\rho$. That is, the lower the $\rho$, the lower and slower will be the decrease in per-capita capital in an economy in which, without migration, the savings rate and its conversion into new capital is insufficient to counteract the depreciation of capital and growth of the population.

\begin{figure}[ht]
\centering
\includegraphics[width=0.6\textwidth]{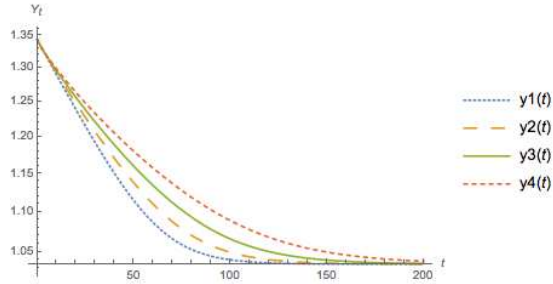}
\caption{The value of per-capita product $y1(t)$ without integer order migration $\rho=1.0$, $y2(t)$ with fractional order of $\rho=0.95$, $y3(t)$ with fractional order of $\rho=0.90$ and $y4(t)$ with fractional order of $\rho=0.85$.}
\label{Fig2}
\end{figure}

Figure (\ref{Fig2}) graphically describes the trajectories of per-capita production without migration and for different values of $\rho>max[\alpha,1-\alpha]$ compared with the $\rho$ of integer order. Figure 2 shows that the per-capita production under the KGCD follows a similar behaviour to that of the per-capita capital. This means, the smaller the parameter $\rho$ is, the lower and slower its trajectory towards the same steady state will be. This is consistent with what happens in the SSM.

\subsection{Case with negative migration}

\begin{figure}[ht]
\centering
\includegraphics[width=0.6\textwidth]{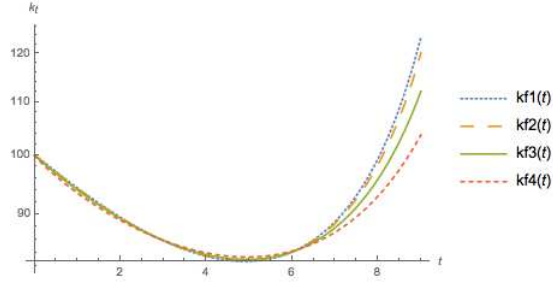}
\caption{The value of the per-capita capital $kf1(t)$ with migration $I=-19.0$ of integer order for $\rho=1.0$, $kf2(t)$ with fractional order of $\rho=0.98$, $kf3(t)$ with fractional order of  $\rho=0.95$ and $kf4$ with fractional order of $\rho=0.90$.}
\label{Fig3}
\end{figure}

Figure (\ref{Fig3}) shows the trajectories of the per-capita capital with negative migration of $-19$ for different values of $\rho>max[\alpha,1-\alpha]$ compared to the one of integer order $(\rho=1)$. From this we can deduce that negative migration generates two different phases in the trajectory of per-capita capital over time: one descending and another ascending. In both phases it happens that, inside of the aforementioned interval, the smaller the $\rho$ value the slower or less fast the decrease and the rise of the per-capita capital trajectories.

\begin{figure}[ht]
\centering
\includegraphics[width=0.6\textwidth]{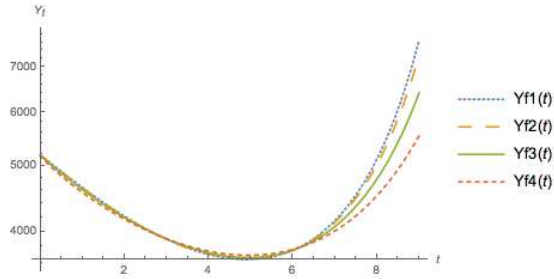}
\caption{The value of per-capita product $yf1(t)$ with migration $I=-19.0$  of integer order for $\rho=1.0$, $yf2(t)$ with fractional order of  $\rho=0.98$, $yf3(t)$ with fractional order of $\rho=0.95$ and $yf4$ with fractional order of $\rho=0.90$.}
\label{Fig4}
\end{figure}

An interesting aspect on this graph is that the lower the value of said para-meter, the corresponding trajectory reaches its vertical asymptote more slowly. This implies that, in the presence of negative migration, if it is possible that per-capita capital, and therefore the per-capita production, will grow indefinitely in a given time $(t^{*})$. The figure (\ref{Fig4}) that appears below shows this. This forces to intuit the divergence in contrast to the convergence predicted by the classical SSM.

Figures \ref{Fig3} and \ref{Fig4} show the possibility of solving the first criticism for the SSM cited in the introduction of this paper. This is provided that three conditions are met: a negative migration $(I<0)$, $I \in (-\infty, -\gamma L_{0})$ and $t<t_{f}$. This is due to the fact that capital and per capita production grow, approaching their vertical asymptote, without stagnation.

\section{Conclusions}

Applying the KGCD to the SSM, we verify the consistency of the model proposed
 with the Inada conditions, which is fulfilled if the $\rho$ value in the KGCD is greater than the maximum of $max[\alpha, 1-\alpha]$, and we obtain closed solutions for capital and per-capita production in cases without migration and with negative migration. The obtained solutions were similar and consistent with the original model. As consequences, we have a simpler model than those of fractional order and time scales, which allows giving a range of possible interpretations to the parameter $\rho$ in terms of economic science. In particular, this model could, without increasing state variables, provide an alternative to solve the first and second criticisms raised in the introduction to classical SSM for the case with negative migration.

For the case without migration as for the classic SSM, with the KGCD it was obtained that at different values of $\rho$ there is convergence of the different trajectories, although at different speeds according to the value of said para-meter.

For the case with negative migration, unlike the classical SSM, it was obtained that to guarantee the convergence of the hypergeometric functions and fulfill the hypothesis that the labour force $L(t)$ must be positive, there is a vali-dity interval for the migration and a finite escape time in which per-capita capital and per-capita production reach a vertical asymptote, without stagnation.

Possible future work with KGCD may generate approximations about the apparent memory effect that has been addressed with fractional derivatives for growth models. In addition, the empirical verification of the convergence proposed of the different trajectories for different economies of the world.

\end{document}